\documentclass[aps,prd,fleqn,superscriptaddress]{revtex4}
\usepackage{graphicx,color,natbib,float}
\usepackage{amsmath,amssymb,amsfonts}
\newcommand{\bse}{\begin{subequations}}
\newcommand{\ese}{\end{subequations}}
\newcommand{\be}{\begin{equation}}
\newcommand{\ee}{\end{equation}}
\newcommand{\bea}{\begin{eqnarray}}
\newcommand{\eea}{\end{eqnarray}}
\newcommand{\ba}{\begin{array}}
\newcommand{\ea}{\end{array}}

\input amssym.def
\input amssym.tex

\usepackage[colorlinks=true, linkcolor=blue, bookmarks=true]{hyperref}
\begin{document}
\hfill%
\vbox{
\halign{#\hfil \cr
IPM/P-2018/015\cr
}
}
\vspace{1cm}
\title{Holographic Mutual and Tripartite Information in a Symmetry Breaking Quench }
\author{M. Asadi}
\email{{\rm{m}}$_{}$ asadi@sbu.ac.ir}
\affiliation{Department of Physics, Shahid Beheshti University G.C., Evin, Tehran 19839, Iran}
\affiliation{School of Physics, Institute for Research in Fundamental Sciences (IPM),
P.O.Box 19395-5531, Tehran, Iran}
\author{M. Ali-Akbari}
\email{{\rm{m}}$_{}$ aliakbari@sbu.ac.ir}
\affiliation{Department of Physics, Shahid Beheshti University G.C., Evin, Tehran 19839, Iran}
\begin{abstract}
We study the time evolution of holographic mutual and tripartite information for a zero temperature $CFT$, derives to a non-relativistic thermal Lifshitz field theory by a quantum quench. We observe that the symmetry breaking does not play any role in the phase space, phase of parameters of sub-systems, and the length of disentangling transition. Nevertheless, mutual and tripartite information indeed depend on the rate of symmetry breaking. We also find that for large enough values of $\delta t$ the quantity $t_{eq}\delta t^{-1}$, where $\delta t$ and $t_{eq}$ are injection time and equilibration time respectively, behaves universally, $i.e.$ its value is independent of length of separation between  sub-systems. We also show that tripartite information is always non-positive during the process indicates that mutual information is monogamous.  
\end{abstract}

\maketitle

\tableofcontents
\section{Introduction and results}
The gauge/gravity duality \cite{Maldacena,Witten}, as the most concrete realization of holographic principle, has been of interest to physicist over the years \cite{Dine,Gross,Witten,Shuryak,Mojaza}. The basic idea is that a gravitational theory defined on a $d+1$ dimensional background, the bulk, is equivalent to a gauge theory defined on a $d$ dimensional spacetime that forms the bulk’s boundary. This correspondence is also a weak/strong duality which has been a useful and powerful tool to study the strongly coupled field theories by gravitational description \cite{Maldacena,Gubser,Witten,Aharony}. Surprisingly, it is extended to the time dependent cases and therefore is appropriate to study the non-equilibrium phenomenon. Various areas raging from Relativistic Heavy Ion Collider to condensed matter physics are tried to explain with this duality. (for a review see \cite{Sachdev,Kovchegov,Gelis,Muller}).\\
Entanglement entropy is one of the most intriguing non-local quantities which measures the quantum entanglement between two sub-systems of a given system. It can be also used to classify the various quantum phase transitions and critical points \cite{Ali-Akbari,Klebanov,Vidal}. Since the quantum field theories have infinitely degrees of freedom, the entanglement entropy is divergent. Thus, it is scheme-dependent quantity and needs to be regulated. It has been shown that the leading divergence term is proportional to the area of the entangleng surface (for $d>2$) \cite{Bombelli,Srednicki}
\begin{eqnarray}
S_{EE}\propto \frac{Area}{\epsilon ^{d-2}},
\end{eqnarray}
where $\epsilon$ is the $UV$ cut-off in quantum field theories. This is called the area law (see also \cite{Casini1,Das}). Note that cut-off dependence of the entanglement entropy makes it to be a non-universal quantity. \\
Due to the $UV$ divergence structure of entanglement entropy, it is natural to introduce an appropriate quantity called mutual information which is an important concept in information theory and has more advantages than the entanglement entropy. It is a finite, positive, semi-definite quantity which measures the total correlation between the two sub-systems $A$ and $B$ \cite{Fischler}. The tripartite information is another useful quantity in this context which is defined for a system consisting of three spatial regions and measures the extensivity of the mutual information. It is also free of divergence and can take any value depending on the underlying field theory. In spite of the mutual information, tripartite information is finite even when the regions share boundaries \cite{Balasubramanian}.\\
To understand $AdS/CFT$ ( for a review see \cite{Aharony}), a particular case of gauge/gravity duality where the gravity lives in a background with a negative cosmological constant, it seems highly important to study how the information in the $CFT$ is encoded in the gravity theory. Since the amount of the information of a sub-system $A$ can be measured by the entanglement entropy of that sub-system, it seems natural to ask how one can calculate this in the gravity side. In \cite{Ryu1,Ryu2}, by applying $AdS/CFT$ correspondence, the authors showed that the entanglement entropy of a region $A$ in a $CFT$ is proportional to the area of a surface which has the minimum area among surfaces whose boundaries coincide with the boundary of the region $A$ which is known as Ryu-Takayanagi ($RT$) prescription. Since both mutual information and tripartite information are combinations of entanglement entropy they can be then calculated by $RT$ prescription. Consequently, If one would like to calculate the amount of the correlation between two sub-system $A$ and $B$, then mutual information is the quantity needs to be computed and the tripartite information is a quantity to study the degree of the extensivity of the mutual information.\\
In this paper we study the time evolution of the holographic mutual and tripartite information of a strongly coupled $CFT$, initial state, which is derived to a non-relativistic fixed point with Lifshitz scaling, final state, by a quantum quench as time evolves. On the gravity side, this non-equilibrium dynamics is equivalent to a background interpolating between a pure $AdS$ at past infinity and an asymptotically Lifshitz black hole at future infinity. We find the following interesting results corresponding to the mutual and tripartite information of the underlying background.
\begin{itemize}
\item The non-equilibrium dynamics following the breaking of the relativistic scaling symmetry leads to the more correlation between two sub-systems. Namely, the less symmetry, the greater correlation.
\item For slow quenches the mutual information approaches the adiabatic regime in the final state, $i.e.$ there is no dependence on the separation length between two sub-systems.
\item Mutual information does undergo a disentangling transition, for a given value of the separation length between two sub-systems, beyond which it is identically zero. Moreover, the separation length of disentangling transition corresponding to the final state is bigger than that of the initial state.
\item There is a specific regime of the parameters, small enough of the length of two sub-systems and their separation length, in the phase space diagram of two sub-systems where the mutual information is independent of the time evolution. 
\item Tripartite information is always non-positive during symmetry breaking quench.
\end{itemize}
\section{Review on background}
The gauge/gravity duality \cite{Maldacena,Witten} provides a wide range of domain to study strongly coupled quantum field theories whose dual are the gravitational theories in one higher dimension. This conjectured duality has been used to explore applications in condensed matter physics and quantum chromodynamics (for a review see \cite{Adams}). In the context of condensed matter, there are quantum systems exhibiting a non-relativistic scaling, which refers to as Lifshitz scaling in the literature, of the following form in $d+1$ dimensions
\begin{eqnarray}
(t,x)\longrightarrow(\lambda^{z} t,\lambda x^{i}),\label{scale}
\end{eqnarray}
where $z$ is a dynamical critical exponent governing the anisotropy between spatial and temporal scaling and $x^{i}$($i=1,2,....d$) denotes the spatial coordinates.
The gauge/gravity logic suggests that one can look for a background metric in one higher dimension than the field theory whose symmetries match with a field theory living on the boundary. In our case the following Lifshitz geometry was proposed in \cite{Hartnoll,Kachru} as a candidate background for the holographic dual of such a non-relativistic theory
\begin{eqnarray}\label{Lif}
ds^{2}= - \frac{r^{2z}}{L^{2z}}dt^{2}+\frac{r^2}{L^2}d\textbf{x}^2+\frac{r^2}{L^2}dr^{2},\label{metric}
\end{eqnarray}
where $l_{AdS}\equiv L(z=1)$, $z$ can take any positive number and the scale transformation acts as (\ref{scale}) along with $r\rightarrow r\lambda^{-1}$. This metric enjoys nice properties such that $(i)$ it is nonsingular and $(ii)$ all local invariants constructed from the Riemann tensor
are constant and finite everywhere \cite{Camilo}.
The case $z = 1$ is the famous Anti-de Sitter spacetime whose symmetry, and its dual scale-invariant theory, is substantially enhanced. $AdS$ geometry is a vacuum solution to a simple $d+1$ dimensional theory of gravity, namely general relativity with a negative cosmological constant
\begin{eqnarray}
S=\frac{1}{16\pi G_{d+1}}\int d^{d+1}x \sqrt{-g}(R+\frac{d(d-1)}{L^{2}}),\label{action1}
\end{eqnarray}
where $G_{d+1}$ is Newton constant and $R$ is the Ricci scalar. Solutions with Lifshitz isometries were first presented in \cite{Kachru}. Einstein gravity with a negative cosmological constant alone does not support the geometry and hence general relativity must be coupled with some matter content. There are many models have been proposed in the literature to reach this Lifshitz solution such as, Einstein-Proca, Einstein-Maxwel-Dilaton and Einstein-$p$ form actions \cite{Kachru,Pang,Taylor,Camilo} or using the nonrelativistic gravity theory of Ho{r}ava-Lifshitz \cite{Griffin}. Here we consider a model involving gravity with negative cosmological constant and a massive gauge field whose action has the following form \cite{Camilo}
\begin{eqnarray}\label{action2}
S=\frac{1}{16\pi G_{d+1}}\int d^{d+1}x \sqrt{-g}[R+d(d-1)-\frac{1}{4}F^{\mu\nu}F_{\mu\nu}-\frac{1}{2} M^{2}A^{\mu}A_{\mu}],
\end{eqnarray} 
where $F^{\mu\nu}$ is the rescaled field strength, corresponding to the rescaled massive gauge field $A^{\mu}$ whose mass is $M$ (For more detailed see \cite{Korovin}).
The Einstein-Proca equations of motion for metric and gauge field are respectively given by
\begin{subequations}
\begin{align}
&R_{\mu \nu}=-d g_{\mu \nu}+\frac{M^2}{2}A_{\mu} A_{\nu}+\frac{1}{2}F_{\mu}^{\sigma}F_{\nu \sigma}+\frac{1}{4(1-d)}F^{\rho \sigma}F_{\rho \sigma}g_{\mu \nu}\, ,\\
&\nabla _{\mu}F^{\mu \nu}=M^{2}A^{\nu}.
\end{align}
\end{subequations}
If one defines 
\begin{eqnarray}
M^{2}=\frac{zd(d-1)^2}{z^2+z(d-2)+(d-1)^2}\, , \qquad L^{2}=\frac{z^{2}+z(d-2)+(d-1)^{2}}{d(d-1)},
\end{eqnarray}
a solution with Lifshitz scaling symmetry can be obtained from action \eqref{action2} 
\begin{subequations}
\begin{align}
&\label{static} ds^{2}= - \frac{r^{2z}}{L^{2z}}dt^{2}+\frac{r^2}{L^2}d\textbf{x}^2+\frac{r^2}{L^2}dr^{2}\, ,\\
& A=\sqrt{\frac{2(z-1)}{z}}\frac{r^{z}}{L^{z}} dt\, ,
\end{align}
\end{subequations}
where the lifshitz scaling can be understood by the following transformation
\begin{eqnarray}\label{Lifshitzscaling}
t\longrightarrow \lambda^{z} t\, , \,\,\,\,\,\,\,x^{i} \longrightarrow \lambda x^{i}\, , \,\,\,\,\,\,\, r\longrightarrow \lambda^{-1}r\, .
\end{eqnarray} 
It is obvious that when $z = 1$ the above solution reduces to the famous $AdS_{d+1}$ solution with unit curvature radius $l_{AdS}=1$.\\
The standard $AdS/CFT$ dictionary states that the presence of the massive gauge field $A_{\mu}$ in the bulk is dual to a vector primary operator $\zeta^{a}$($a=0,1,...d$) of dimension $\Delta$ \cite{Korovin}
\begin{eqnarray}
\Delta=\frac{1}{2}[d+\sqrt{(d-2)^2+4M^2}]=\frac{d}{2}+\sqrt{\frac{(d-2)^2}{4}+\frac{zd(d-1)^2}{z^2+z(d-2)+(d-1)^2}}\, .
\end{eqnarray}
In other words, one can say that the action (\ref{action2}) controls the dynamics of a $CFT$ whose spectrum contains a vector primary operator of dimension $\Delta$. The asymptotic expansion of the bulk gauge field is also given by
\begin{eqnarray}
A_{t}=r^{\Delta -d +1} A_{t}^{(0)}+....... +\,r^{-(\Delta -1)} A_{t}^{(d)}+......\, ,
\end{eqnarray}
where $A_{t}^{(0)}$ is the source of the dual operator and $A_{t}^{(d)}$ is related to its expectation value.
It was shown in \cite{Korovin} that the Lifshitz geometries get close to $AdS $ when dynamical exponent $z$ is close to unity, $i.e.$ $z=1+\epsilon^{2}$
where $\epsilon\ll1$. In this case the static solution (\ref{static}) reads
\begin{subequations}
\begin{align}
&ds^2=-r^2[1+2\epsilon^2\ln r+\frac{\epsilon^2}{1-d}]dt^{2}+r^{2}[1+\frac{\epsilon^2}{1-d}]d\textbf{x}^2+[1-\frac{\epsilon^2}{1-d}]\frac{dr^2}{r^2}+O(\epsilon^4)\, , \\ 
&\label{gauge1} A=\sqrt{2}\epsilon r dt + O(\epsilon ^{3}) \,,
\end{align}
\end{subequations}
and the corresponding mass $M$ and the dual operator dimension $\Delta$ have also the following expressions
\begin{eqnarray}
M^{2}=d-1+(d-2)\epsilon^{2}+O(\epsilon^{4}) \, ,\qquad\Delta=d+\frac{d-2}{d}\epsilon^{2}+O(\epsilon^{4}) \,.
\end{eqnarray}
In this case the asymptotic expansion of the bulk gauge field is given by
\begin{eqnarray}\label{gauge2}
A_{t}=r(1+O(\epsilon ^{2})) A_{t}^{(0)}+.......+\, r^{-(d -1)}(1+O(\epsilon ^{2})) A_{t}^{(d)}+......\, ,
\end{eqnarray}
According to (\ref{gauge1}) and (\ref{gauge2}) and by the following identifications
\begin{eqnarray}
A_{t}^{(0)}\equiv \sqrt{2}\epsilon + O(\epsilon ^{3})\, , \qquad A_{t}^{(d)}\equiv O(\epsilon^{3})\, , \qquad \Delta=d \, ,
\end{eqnarray}
one can extend the standard $AdS/CFT$ dictionary in order to study the dual field theory. Indeed, this special class of Lifshitz spacetime can be considered holographically as a continuous deformation of corresponding $CFT$ by time component of a vector primary operator $\zeta^{a}$ of conformal dimension $\Delta=d$, namely
\begin{equation}\label{action lif}
S_{Lif}=S_{CFT}+\sqrt{2}\epsilon\int d^{ d}x \zeta^{t}(x).
\end{equation}
It is worth to mention that many Lifshitz invariant solutions exist which are not of the above form and then, holographically, one can not reach the key features of dual theory.\\
In \cite{Camilo} the authors consider a nice mechanism to study the symmetry breaking of a $CFT$ towards a non-relativistic Lifshitz scaling with $z=1+\epsilon^{2}$. In fact, they consider a quantum quench profile $j(t)\equiv \sqrt{2}\epsilon J(t)$, coupled to the vector primary operator $\zeta^{t}(x)$ in the action (\ref{action lif}) which interpolates smoothly between $0$ and $\sqrt{2}\epsilon$. The first corresponds to a strongly coupled $CFT$ at zero temperature (initial state) and the later to a finite temperature fixed point with Lifshitz scaling (thermal finite state) (\ref{Lifshitzscaling}) as time evolves from past infinity to future infinity. The new action governing this process has the following form
\begin{eqnarray}
S=S_{CFT}+\sqrt{2}\epsilon\int d^{ d}x J(t) \zeta^{t}(x).
\end{eqnarray}
In the following we merely review the above mechanism which has already done in \cite{Camilo}. Working with the ingoing Eddingtone-Finkelstein($EF$) coordinate system ($\nu , r , \textbf{x}$) and arbitrary exponent $z$, consider the following ansatz for the metric and the gauge field 
\begin{subequations}\label{ansatz}
\begin{align}
&ds^{2}=2h(\nu ,r)d\nu dr-f(\nu ,r)d\nu ^{2}+r^{2}d\textbf{x}^2 \, , \\
&A(\nu ,r)=a(\nu ,r)d\nu +b(\nu ,r)dr \, ,
\end{align}
\end{subequations}
where $h,f,a$ and $b$ are four unknown functions. In order to focus on the case of interest, $i.e.$ $z=1+\epsilon ^{2}$, one should expand $h(\nu ,r),f(\nu ,r),a(\nu ,r)$ and $b(\nu ,r)$ in the ansatz (\ref{ansatz}) as a power series in $\epsilon$, that is
\begin{subequations}
\begin{align}
&f(\nu ,r)=\sum _{n=0}^{\infty}f^{n}(\nu ,r)\epsilon^{n} \, ,\\
&h(\nu ,r)=\sum _{n=0}^{\infty}h^{n}(\nu ,r)\epsilon^{n} \, ,\\
&a(\nu ,r)=\sum _{n=0}^{\infty}a^{n}(\nu ,r)\epsilon^{n} \, ,\\
&b(\nu ,r)=\sum _{n=0}^{\infty}b^{n}(\nu ,r)\epsilon^{n} \, ,
\end{align}
\end{subequations}
and then solves the equations of motion corresponding to the action (\ref{action2}) in terms of $\epsilon$ expansion, to leading non-trivial order for each function (in this case up to $\epsilon ^{2}$). To solve the equations of motion for a given order in $\epsilon$, consider the following ansatz for $f^{n}(\nu ,r),h^{n}(\nu ,r),a^{n}(\nu ,r)$ and $b^{n}(\nu ,r)$
\begin{subequations}
\begin{align}
&f^{n}(\nu ,r)=r^{2}\sum _{l=0}(f_{l}^{(n)}(\nu)+\tilde{f}_{l}^{(n)}(\nu)\ln r)\, r^{-l} \, , \\
&h^{n}(\nu ,r)=\sum _{l=0}(h_{l}^{(n)}(\nu)+\tilde{h}_{l}^(n)(\nu)\ln r)\,r^{-l} \, , \\
&a^{n}(\nu ,r)=r\sum _{l=0}(a_{l}^{(n)}(\nu)+\tilde{a}_{l}^{(n)}(\nu)\ln r)\,r^{l} \, , \\
&b^{n}(\nu ,r)=\frac{1}{r}\sum _{l=0}(f_{l}^{(n)}(\nu)+\tilde{f}_{l}^{(n)}(\nu)\ln r)\,r^{l} \, ,
\end{align}
\end{subequations}
along with the initial conditions (note that in the boundary side the initial state corresponds to a zero temperature state of the strongly coupled $CFT$ which is represented by a pure $AdS$ geometry in the bulk with no gauge field)
\begin{subequations}\label{initial con}
\begin{align}
&f(\nu\rightarrow -\infty ,r)=r^{2}\, ,\\
&h(\nu\rightarrow -\infty ,r)=1\, ,\\
&a(\nu\rightarrow -\infty ,r)=0\, ,\\
&b(\nu\rightarrow -\infty ,r)=0\, ,
\end{align}
\end{subequations}
and the boundary conditions at $r\rightarrow\infty$ to order $\epsilon^{2}$ 
\begin{subequations}\label{boundary con}
\begin{align}
&f(\nu ,r\rightarrow \infty)=r^{2}(1+2\epsilon^{2}J(\nu)^{2}\ln r+.....)\, ,\\
&h(\nu ,r\rightarrow \infty)=1+2\epsilon^{2}J(\nu)^{2}\ln r+.....)\, ,\\
&a(\nu ,r\rightarrow \infty)=\sqrt{2}\epsilon J(\nu) r+.....\, ,\\
&a(\nu ,r\rightarrow \infty)=0 \, ,
\end{align}
\end{subequations}
where $J(\nu)$ is the quench profile which specifies how energy is injected in to the system. Considering a quantum quench of a $CFT$, living in ($2+1$) dimensions, and following the underlying scheme along with the initial conditions (\ref{initial con}) and the boundary conditions (\ref{boundary con}) the solution for the metric and gauge field to order $\epsilon^{2}$ reads \cite{Camilo}
\begin{subequations}
\begin{align}
&\label{final form1} A(\nu ,r)=\epsilon[a^{(1)}(\nu ,r)d\nu +b^{(1)}(\nu ,r)dr]+O(\epsilon^{3})\, ,\\
&\label{final form2} ds^{2}=2[1+\epsilon^{2}h^{(2)}(\nu ,r)]d\nu dr-[r^{2}+\epsilon^{2}f^{(2)}(\nu ,r)]d\nu ^{2}+r^{2}(dx_{1}^{2}+dx_{2}^{2})+O(\epsilon^{4}) \, ,
\end{align}
\end{subequations}
where $a^{(1)}(\nu ,r),b^{(1)}(\nu ,r),f^{(2)}(\nu ,r)$ and $h^{(2)}(\nu ,r)$ are given by
\begin{subequations}
\begin{align}
&a^{(1)}(\nu ,r)=\sqrt{2}r(J(\nu)+\frac{\dot{J}(\nu)}{r}+\frac{\ddot{J}(\nu)}{2r^{2}})\, ,\\
&b^{(1)}(\nu ,r)=\frac{-\sqrt{2}}{r}(J(\nu)+\frac{\dot{J}(\nu)}{2r})\, ,\\
&f^{(2)}(\nu ,r)=2r^{2}(\ln r-\frac{1}{4})J(\nu)^{2}-3rJ(\nu)\ddot{J}(\nu)-\ddot{J}(\nu)^{2}-\frac{I(\nu)}{r}\, ,\\
&h^{(2)}(\nu ,r)=J(\nu)^{2}\ln r-\frac{J(\nu)\ddot{J}(\nu)}{r}-\frac{\ddot{J}(\nu)^{2}}{8r^{2}} \, , 
\end{align}
\end{subequations}
and the coefficient $I(\nu)$ is defined 
\begin{eqnarray}
I(\nu)=\frac{1}{2}\int _{-\infty}^{\nu} \ddot{J}(\omega)^{2}d\omega \, .
\end{eqnarray}
In the limit $\nu\rightarrow-\infty$, for which $J(\nu)$ goes to zero, we are left with the static $AdS$ solution with no gauge field (zero temperature initial state) and in the limit $\nu\rightarrow\infty$, for which $J(\nu)$ goes to one, the final state corresponds to an asymptotically Lifshitz black brane (thermal finial state) as follows
\begin{eqnarray}
ds_{f}^{2}=2(1+\epsilon^{2}\ln r)d\nu dr-r^{2}[1+2\epsilon^{2}(\ln r-\frac{1}{4})-\epsilon^{2}\frac{I_{f}}{r^{3}}]d\nu ^{2}+r^{2}(dx_{1}^{2}+dx_{2}^{2})+O(\epsilon^{4}) \, ,
\end{eqnarray}
whose event horizon will be located at $r=r_{h}$ given by the largest solution of the following equation
\begin{eqnarray}\label{horizon1}
1+2\epsilon^{2}(\ln r_{h}-\frac{1}{4})-\epsilon^{2}\frac{I_{f}}{r_{h}^{3}}=0 \, .
\end{eqnarray}
In \cite{Camilo} two specific quench profiles, as a probe of the quench dynamic, have been considered to study both local observable such as vacuum expectation values of the stress-energy tensor and of the quenching operator and also non-local one such as entanglement entropy. However, in this paper, we concentrate on the following profile
\begin{eqnarray}
J(\nu)=\frac{1}{2}(1+\tanh \frac{\nu}{\delta t}),
\end{eqnarray}
where $\delta t$ is a time scale which we call it the quenching time. At the asymptotic boundary $r=\infty$ both $\nu$ and $t$ coincide, thus, one can understand the bulk quench profile $J(\nu)$ as $J(t)$ for an observer living on the boundary side.
The discussed mechanism is merely valid for values of $r\rightarrow\infty$ (boundary) up to $r\sim r_{h}$ where the event horizon of the final state black hole obtained from (\ref{horizon1}) and given by \cite{Camilo}
\begin{eqnarray}\label{horizon2}
r_{h}\simeq \frac{0.5 \epsilon ^{\frac{2}{3}}}{\delta t}.
\end{eqnarray}
It is noticed that the temperature of the Lifshitz final state, $T$, is also proportional to $r_{h}$ or equivalently
\begin{eqnarray}\label{temperature}
T \propto r_{h}\propto \frac{\epsilon ^{\frac{2}{3}}}{\delta t}. 
\end{eqnarray}
\section{Review on the entanglement entropy, mutual information and tripartite information}
\begin{itemize}
\item \textbf{Entanglement entropy:} The entanglement entropy is one of the most important quantities which measures the quantum entanglement among different degrees of freedom of a quantum mechanical system \cite{Horodecki,Casini}. In fact, entanglement entropy has emerged as a valuable tool to probe the physical information in quantum systems.\\
To define the entanglement entropy we decompose the total system into two sub-systems $A$ and its complement $\bar{A}$. Accordingly, the total Hilbert space $\mathcal{H}$ becomes a direct products of $\mathcal{H}_{A}$ and $\mathcal{H}_{\bar{A}}$ such that
\begin{eqnarray}
\mathcal{H} = \mathcal{H}_{A} \otimes \mathcal{H}_{\bar{A}}.
\end{eqnarray}
We then define the reduced density matrix $\rho_{A}$ for the sub-system $A$ by integrating out the degrees of freedom $\bar{A}$
\begin{eqnarray}
\rho_{A} = Tr _{\bar{A}} [\rho],
\end{eqnarray}
where $\rho$ is the total density matrix of the entire system. Then the entanglement entropy is defined as the Von-Neumann entropy for $\rho_{A}$
\begin{eqnarray}
S_{A} = -Tr [\rho_{A} \log \rho_{A} ].
\end{eqnarray}
When the system is in a pure state the Von-Neumann entropy of the complete system is zero and the following property is fulfilled
\begin{eqnarray}
S_{A} = S_{ \bar{A}}.
\end{eqnarray}
Noticeably, in a quantum field theory, entanglement entropy of a region $A$ contains short-distance divergence and behaves according to an area law \cite{Srednicki}. It can be shown that the entanglement entropy for two disjoint sub-systems $A_{1}$ and $A_{2}$ satisfies the so-called strong subadditivity condition \cite{Lieb}
\begin{eqnarray}\label{subadd}
S(A_{1}) + S(A_{2}) \geq S(A_{1}\cup A_{2})+S(A_{1}\cap A_{2}).
\end{eqnarray}
According to $AdS/CFT$ correspondence, for large $N$ theories on the boundary side there exist gravity dual theories on the bulk side which are described by classical Einstein gravity with suitable matter field content. The holographic entanglement entropy of a sub-system $A$ on the boundary field theory can be computed using the $RT$ prescription proposed in \cite{Ryu1,Ryu2}
\begin{eqnarray}\label{RT}
S_{A}=\frac{Area (\gamma_{A})}{4 G_{d+1}} \, ,
\end{eqnarray}
where $\gamma_{A}$ is the area of the minimal surface, extended in the bulk, whose boundary coincides with the boundary of the $\partial A$ (so that $\partial \gamma_{A} = \partial A $) . We also require that $\gamma_{A}$ is homologous to $A$.
\end{itemize}
\begin{itemize}
\item \textbf{Mutual information:}
Having introduced the entanglement entropy for a sub-system $A$ and its complement $\bar{A}$ if one would like to measure the amount of correlation between two disjoint regions $A$ and $B$, the most interesting finite quantity is the mutual information
\begin{eqnarray}\label{MI}
I(A,B)\equiv S(A)+S(B)-S(A\cup B) \, ,
\end{eqnarray}
where $S(X)$ denotes the entanglement entropy corresponding to the region $X$.
While the entanglement entropy of a region contains $UV$ divergence proportional to the area of the boundary of $A$, the mutual information is free of divergence (finite) since in this linear combination the leading divergence due to the area law is canceled. It is noticed that due to the subadditivity condition (\ref{subadd}) the mutual information indeed satisfies the following inequality \cite{Allais}
\begin{eqnarray}\label{Sub}
I(A,B)\geq 0\, ,
\end{eqnarray}
\begin{figure}[h] 
\centering
\includegraphics[width=0.75\textwidth]{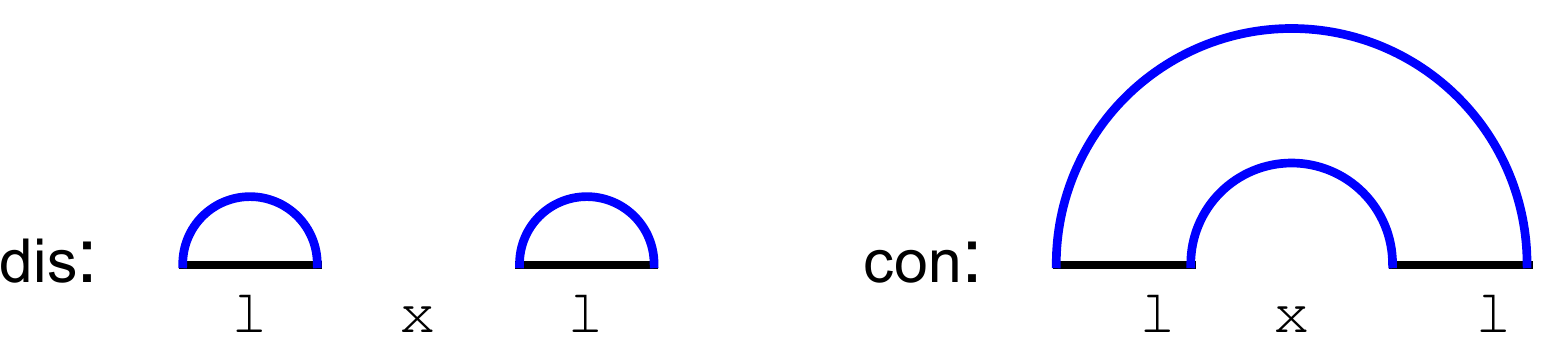} 
\caption{Two different configurations for computing $S_{A\cup B}$. The time coordinate is suppressed.}
\label{fig:s}
\end{figure}
with equality if and only if $A$ and $B$ are uncorrelated. It was pointed out in \cite{Headrick} mutual information indeed undergoes a disentangling transition as one increases the separation between the two sub-systems $A$ and $B$ that is for small separation, $I(A, B) \neq 0$ but $I(A, B) = 0$ for large separation. On the other hand, when $I(A, B) = 0$ there is no correlation between two sub-systems and hence they become completely decoupled \cite{Fischler}.\\
Imagine two disjoint sub-systems $A$ and $B$, of the same length $l$ separated by $x$, the entanglement entropy of each sub-system can be computed from (\ref{RT}). However, the computation of $S(A\cup B)$ is more interesting. In the bulk, depending on the critical ratio $\frac{x}{l}$, there are two candidate minimal surfaces which are schematically shown in Fig. \ref{fig:s} and we therefore have
\begin{eqnarray}\label{sAB}
{S_{A\cup B}}=
\begin{cases}
2 S(l), & \text{large}\: \frac{x}{l}, \\
S(2l+x) +S(x), & \text{small}\: \frac{x}{l}, \\
\end{cases}
\end{eqnarray}
where $S(Y)$ denotes the area of the minimal surface whose boundary is coincided with the boundary of the underlying sub-system. Accordingly, one can immediately reach the following result for the mutual information
\begin{eqnarray}\label{MI}
{I(A,B)}=
\begin{cases}
0, \,\,\,\,\,\,\,\,\,\,\,\,\,\,\,\,\,\,\,\,\,\,\,\,\,\,\,\,\,\,\,\,\,\,\,\,\,\,\,\,\,\,\,\,\,\,\,\,\,\,\,\,\,\,\,\,\,\,\,\,\,\,\,\,\,\,\,\,\,\,\,\,\,\,\text{large}\: \frac{x}{l}, \\
2S(l) - S(2l+x) -S(x), \qquad \text{smal}\: \frac{x}{l}. \\
\end{cases}
\end{eqnarray}
Mutual information can potentially provide a powerful description of how correlations evolve and spread in an out-of-equilibrium system which is of our interest in this paper.
\end{itemize}
\begin{figure}[h] 
\centering
{\includegraphics[width=0.70\textwidth]{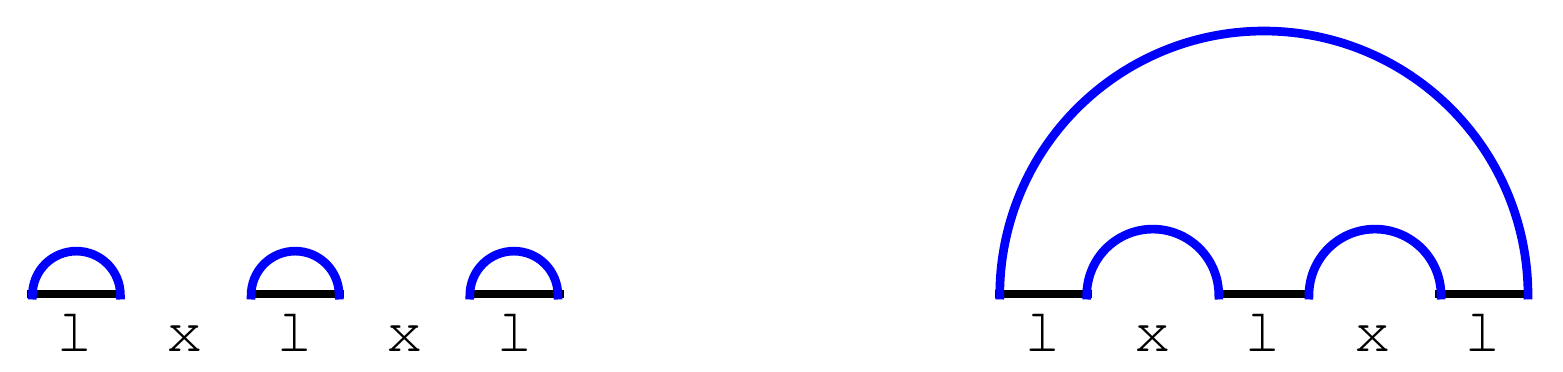} }\\
{\includegraphics[width=0.70\textwidth]{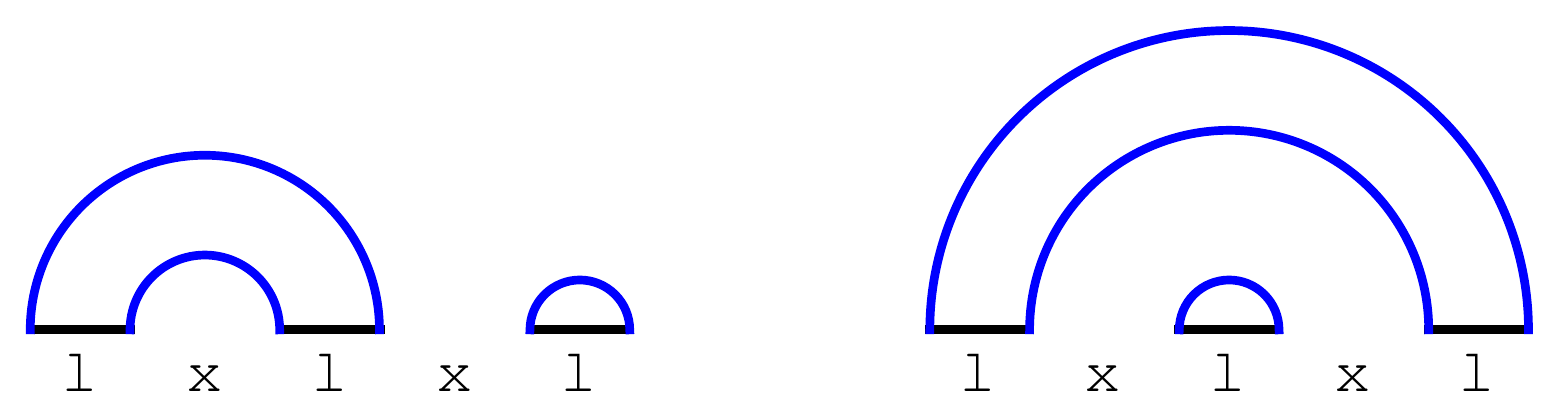} }
\caption{Four different configurations for computing $S_{A\cup B \cup C}$. The time coordinate is suppressed. } \label{figI3}
\end{figure}
\begin{itemize}
\item \textbf{Tripartite information:} In addition to the mutual information there is another interesting quantity, defined from the entanglement entropy, called the tripartite information 
\begin{eqnarray}\label{I3}
I^{[3]}(A\cup B\cup C)\equiv S(A)+S(B)+S(C)-S(A\cup B)-S(A\cup C)-S(B\cup C)+S(A\cup B\cup C)\, ,
\end{eqnarray}
where $A , B$ and $C$ are three disjoint intervals. It is clear that this quantity is symmetric under permutations of its arguments and it is also free of divergences even when the regions share their boundary \cite{Hayden}. According to (\ref{I3}) the tripartite information can be also written in terms of the mutual information as follows
\begin{eqnarray}
I^{[3]}(A\cup B\cup C)\equiv I(A\cup B) + I(A\cup C)- I(A\cup B\cup C)\, . 
\end{eqnarray}
Tripartite information can measure the degree of extensivity of the mutual information in such a way that mutual information is extensive when $I^{[3]} = 0$, superextensive when $I^{[3]} < 0$ and subextensive when$ I^{[3]} > 0$. In either the extensive or the superextensive case mutual information is said to be monogamous. It is noticed that when $I^{[3]} = 0$ the mutual information of $A$ with $BC$ is the sum of its mutual information with $B$ and $C$ individually.
While in a generic quantum system, the tripartite information can be positive, negative or zero, depending on the choice of the regions, but in \cite{Hayden} it was shown that according to the RT prescription the mutual information is always monogamous, $i.e.$
\begin{eqnarray}
I^{[3]}(A \cup B \cup C) \leq 0\, ,
\end{eqnarray}
for any regions $A, B$ and $C$ in the boundary field theory. To calculate the tripartite information for three disjoint regions $A , B$ and $C$ of the same length $l$ with separation $x$ the computation of $S(A\cup B\cup C)$ is more challenging. For the union of three subsystems one should consider different configurations for the extremal surfaces. In fact, considering the $N$ intervals we should compare $(2N-1)!! $ configurations ($N=3\Rightarrow 15 $ configurations in our case). However, one can show that for $N = 3$ equal intervals we are left only with the four configurations depicted in Fig. \ref{figI3} \cite{Allais}.\\
\end{itemize}
In \cite{Camilo} the authors specialized to the $3$-dimensional boundary and considered a strip-like entangling region $A$ with the length $l$ in the $x^{1}$ direction and regulated length $l_{\perp}\rightarrow \infty$ in the $x^{2}$ direction at a constant time slice and they reached the following expression for the entanglement entropy 
\be\begin{split}\label{EE}
\delta s_{A_{finite}}(t)&\equiv s_{A}(t)-s_{A}^{(0)}\\ \,\,\,&=\epsilon^{2}\frac{l_{\perp}}{4G_{3}}\left\{ \int _{r_{\ast}}^{\infty} dr \frac{\sqrt{r^{4}-r_{\ast}^{4}}}{r^{2}}\,\,\,\,\,[\frac{J(t-\frac{1}{r})^{2}-J(t)^{2}}{2}+\frac{J(t-\frac{1}{r})\dot{J}(t-\frac{1}{r})}{r}+\frac{3\dot{J}(t-\frac{1}{r})^{2}}{4r^{2}}+\frac{I(t-\frac{1}{r})}{r^{3}}]\right.\\ &\left.
+\frac{\sqrt{\pi}\Gamma (\frac{-1}{4})}{16\Gamma(\frac{5}{4})}r_{\ast}J(t)^{2}+\frac{5}{2}J(t)\dot{J}(t) \right\},
\end{split}
\ee
where $r_{\ast}$ is related to the boundary size of the entangling region through $r_{\ast}=\frac{1.19814}{l}$ and the time independent background contribution to the entanglement entropy $s_{A}^{(0)}$ has been subtracted to study the time evolution of the entanglement entropy \cite{Camilo}. They also defined $\delta S_{A_{finite}}(t)\equiv \frac{4G_{3}}{l_{\perp}}\delta s_{A_{finite}}(t)$ as a more useful quantity to study the effect of quenching process on the time evolution of the underlying theory. It is worth to recall that one can trust the upcoming calculations by demanding that $r_{\ast}> 2 r_{h}$ which means that the size of the sub-systems must satisfies the following inequality
\begin{eqnarray}\label{condition}
l < \frac{1.19814\, \delta t}{\epsilon^{\frac{2}{3}}}\, ,
\end{eqnarray}
which is sufficiently away from the event horizon of the final state Lifshitz black brane. \\
In the following we consider sub-systems of strip-like shape in the field theory described by background \eqref{final form2} and study the time evolution of the mutual and tripartite information.
\section{Numerical results}
\begin{figure}[h] \label{fig:MI1}
\centering
{\includegraphics[width=0.48\textwidth]{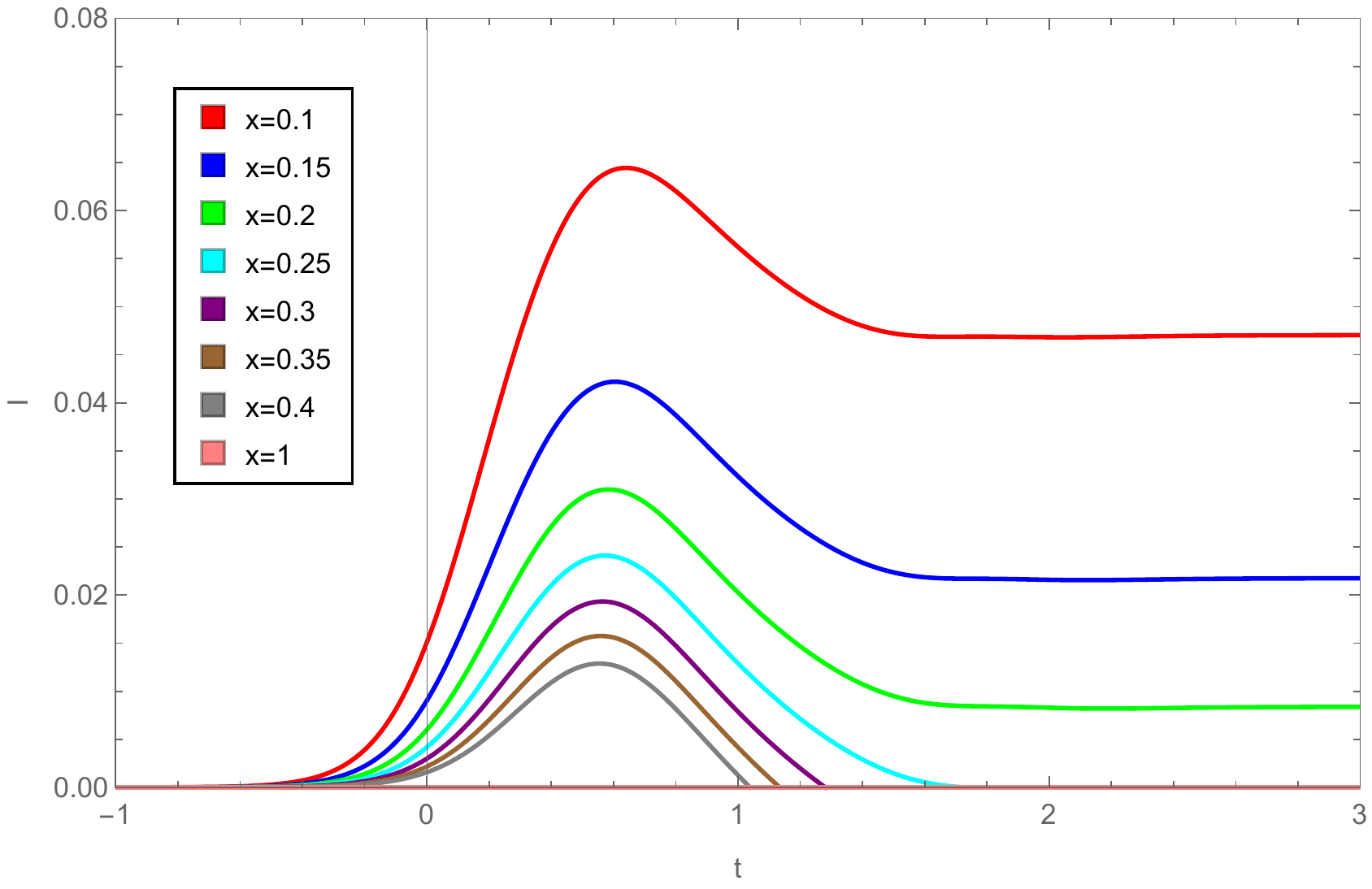} }
{\includegraphics[width=0.48\textwidth]{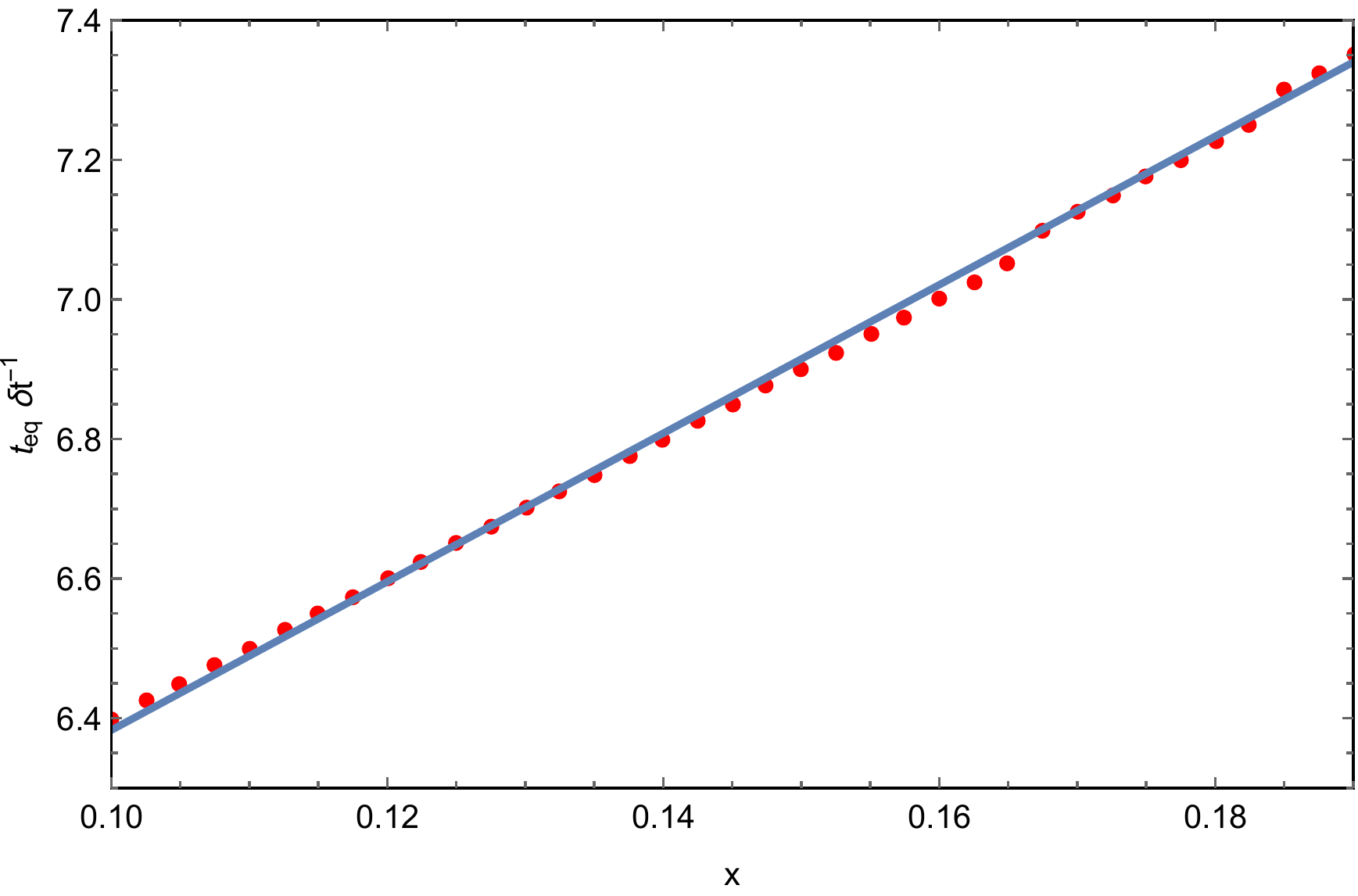} }
\caption{$Left$: The rescaled holographic mutual information $I$ as a function of the boundary time $t$ at fixed $ l=1$ and $\delta t=0.4 $ . The different curves are characterized by different values of $x=0.1$ (top) to $0.4$ (bottom). (Some of the curves are not visible since everywhere vanishing). $Right$ : Rescaled equilibrium time $t_{eq} \delta t ^{-1}$ as a function of separation $x$ at fixed $ l=1$ and $\delta t=0.4 $. The larger separation $x$ , the longer rescaled equilibrium time $t_{eq} \delta t ^{-1}$. }\label{fig1}
\end{figure}
%
In Fig. \ref{fig1} we show the time evolution of the mutual information of two sub-systems, with the same length $l$, in the left plot and the dependence of rescaled equilibration time on the separation length $x$ in the right plot. We set $\epsilon =0.1$ which means that the dynamical exponent of the final state Lifshitz theory will be $z=1.01$.
In the left plot the length of the two sub-systems and the injection time are to be fixed $l=1$ and $\delta t=0.4$, respectively, to study the effect of separation length $x$ on mutual information. It can be seen that by varying the separation $x$ between them three different behaviors occur. Firstly, as it was expected, for very large $x$ mutual information is zero at all times (pink curve $x=1$) . Secondly, for large enough $x$, the mutual information undergoes a transition beyond which it is identically zero. In fact, when $I(A,B)=0$ the two sub-systems $A$ and $B$ become completely decoupled and hence one would say that a disentangling transition occurs. Thirdly, for a given value of $x$, \textit{i.e.} $x=0.25,0.3,0.35,0.4$ in our case, it is clear that the mutual information is (approximately) zero at $t=-\infty$ then it reaches positive values at intermediate times, before the thermal state is reached, and then vanishes at the end . Finally and more interestingly, for small enough $x$ we find that the mutual information is positive for any boundary time $t$ which means that there always exist correlation between the two sub-systems at all times. From another point of view in this case one can say that the connected configuration, see Fig .\ref{fig:s}, would be always minimal.
Consequently, the mutual information does respect to the subadditivity condition (\ref{Sub}). 
These results are in complete agreement with the ones reported in the literature, see $e.g.$ \cite{Allais, Balasubramanian,Fischler,Hayden}. One can also study time evolution of mutual information for different length scale $l$, $x$ and different injection times $\delta t$. The logic is identical and the same results will be obtained.
We would like to define $t_{eq}$ as a specific time above which the mutual information reaches its equilibrium value. To do so, we consider the following time dependent function 
\begin{eqnarray}
\epsilon(t) =\vert\frac{I(t=\infty)-I(t)}{I(t=\infty)}\vert,
\end{eqnarray}
where the equilibration time is defined as the time which staisfies $\varepsilon(t_{eq})<0.001$ and $\varepsilon(t)$ stays below this afterwards. In the right plot of Fig. \ref{fig3} we set $ l=1$ and $\delta t=0.4 $ to analyze relation between the separation length $x$ and the rescaled equilibrium time $t_{eq} \delta t ^{-1}$. It is noticed that we have just considered curves which have $I(A, B)\neq0$ at $t\rightarrow\infty$.
Here we list the following interesting features corresponding to Fig. \ref{fig1}.
\begin{enumerate}
\item \textbf{Left plot:}
\begin{itemize}
\item Interestingly, for large enough $x$ there is a disentangling transition before the two sub-systems reach the final equilibrium. That is when the two sub-systems become largely separated they become decoupled and have no time to reach the equilibrium, at least from the mutual information point of view. 
\item There is always a value of $x$ for which the disentangling transition occurs, either in the initial state or the final state, which we call it $x^{DT}$. Our results indicate that $x^{DT}_{Lif}>x^{DT}_{AdS}$, $i.e.$ in the final state with lower symmetry $x^{DT}$ increases. However, notice that $x^{DT}$ is independent of how much the symmetry is broken, since $\epsilon$ merely shifts the mutual information according to (\ref{EE}).
\item We can observe that small enough separations $x < l$ (top curves) cause the mutual information at late times ($t=+\infty$) be always greater than those at earlier times ($t=-\infty$). In fact, the non-equilibrium dynamics following the breaking of the relativistic scaling symmetry leads to the more correlation between two sub-systems. Namely, the less symmetry, the greater correlation. Note that although the final state is thermal and hence the temperature decreases the mutual information but the above statement is still correct. 
\end{itemize}
\item \textbf{Right plot:}
\begin{itemize}
\item It is obvious from the figure that if we decrease the separation length $x$, the two subsystems reach their final state faster. In fact, the smaller the separation, the faster the rescaled equilibration time.
\item Interestingly, there is a linear relationship between the separation length $x$ and the rescaled equilibrium time $t_{eq} \delta t ^{-1}$ of the following form
\begin{eqnarray}
t_{eq} \delta t ^{-1} = 10.636 x +5.319.
\end{eqnarray}
\item According to (\ref{MI}) and (\ref{EE}) if one would like to break the symmetry of the initial state more strongly, $i.e.$ choosing larger $\epsilon$, the holographic mutual information merely experiences a shift and the rescaled equilibrium time is independent of the rate of symmetry breaking.
\end{itemize}
\end{enumerate}
\begin{figure}[h] \label{fig:MI1}
\centering
{\includegraphics[width=0.48\textwidth]{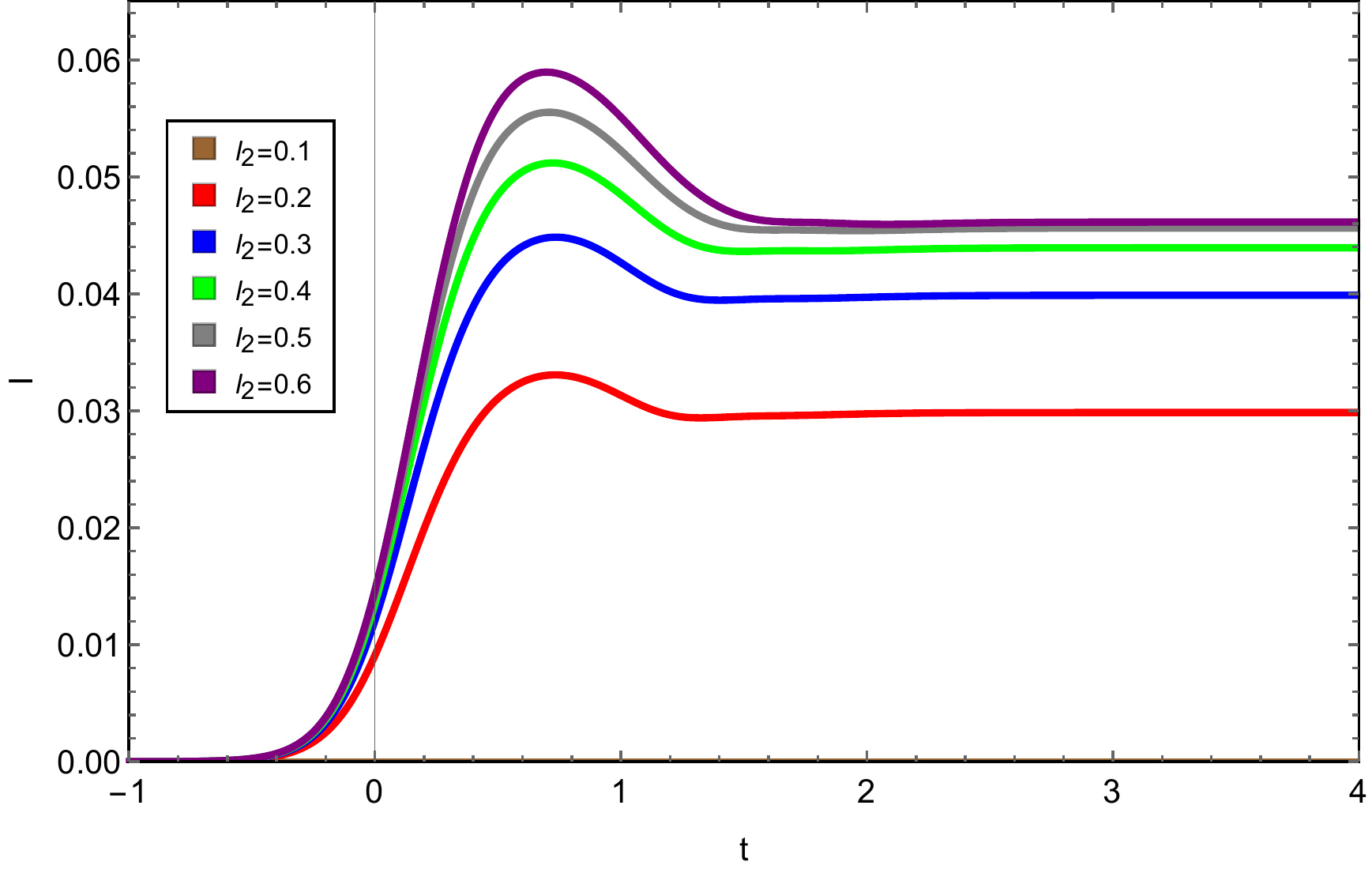} }
{\includegraphics[width=0.48\textwidth]{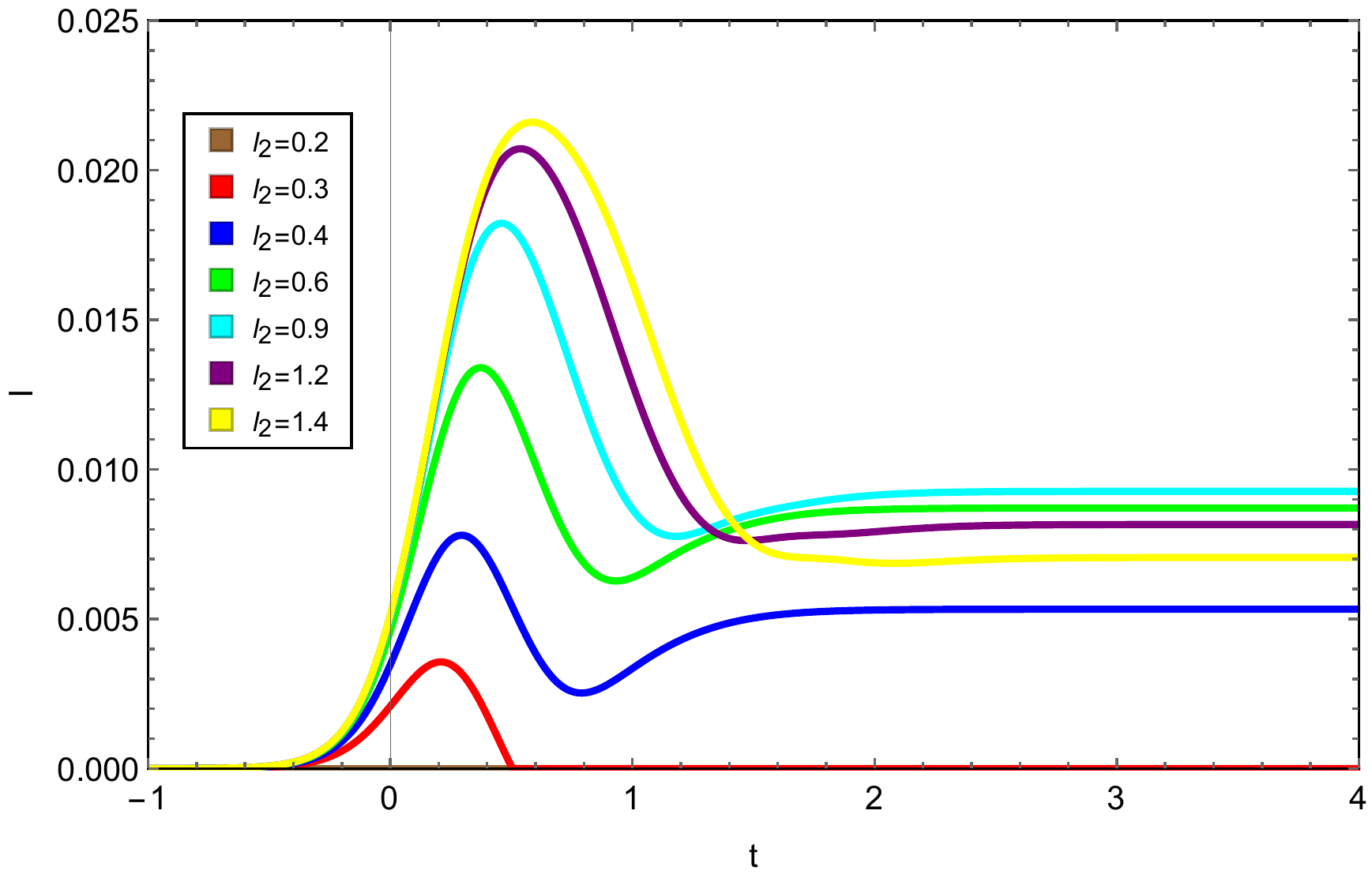} }\\
{\includegraphics[width=0.49\textwidth]{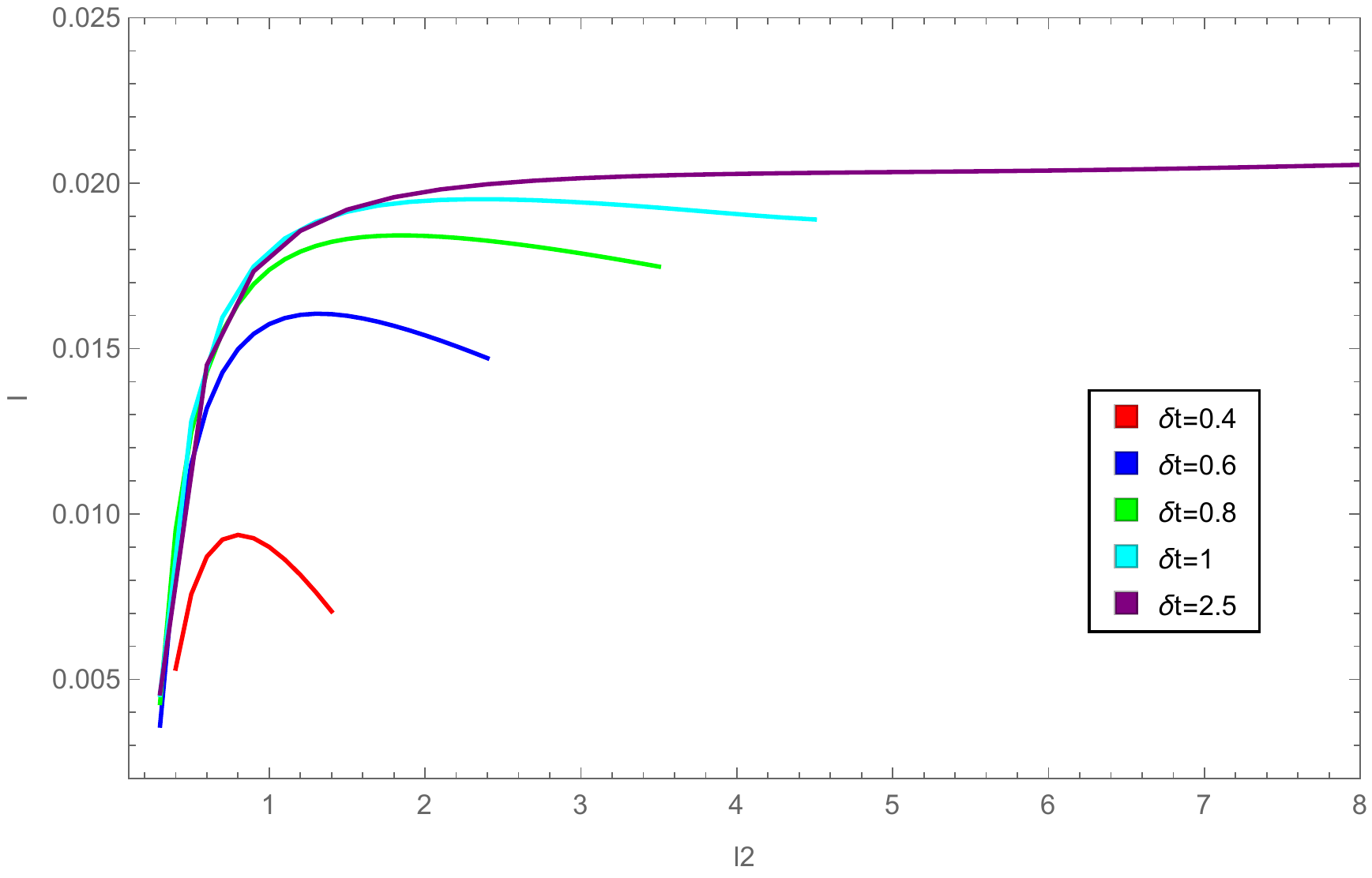} }
{\includegraphics[width=0.49\textwidth]{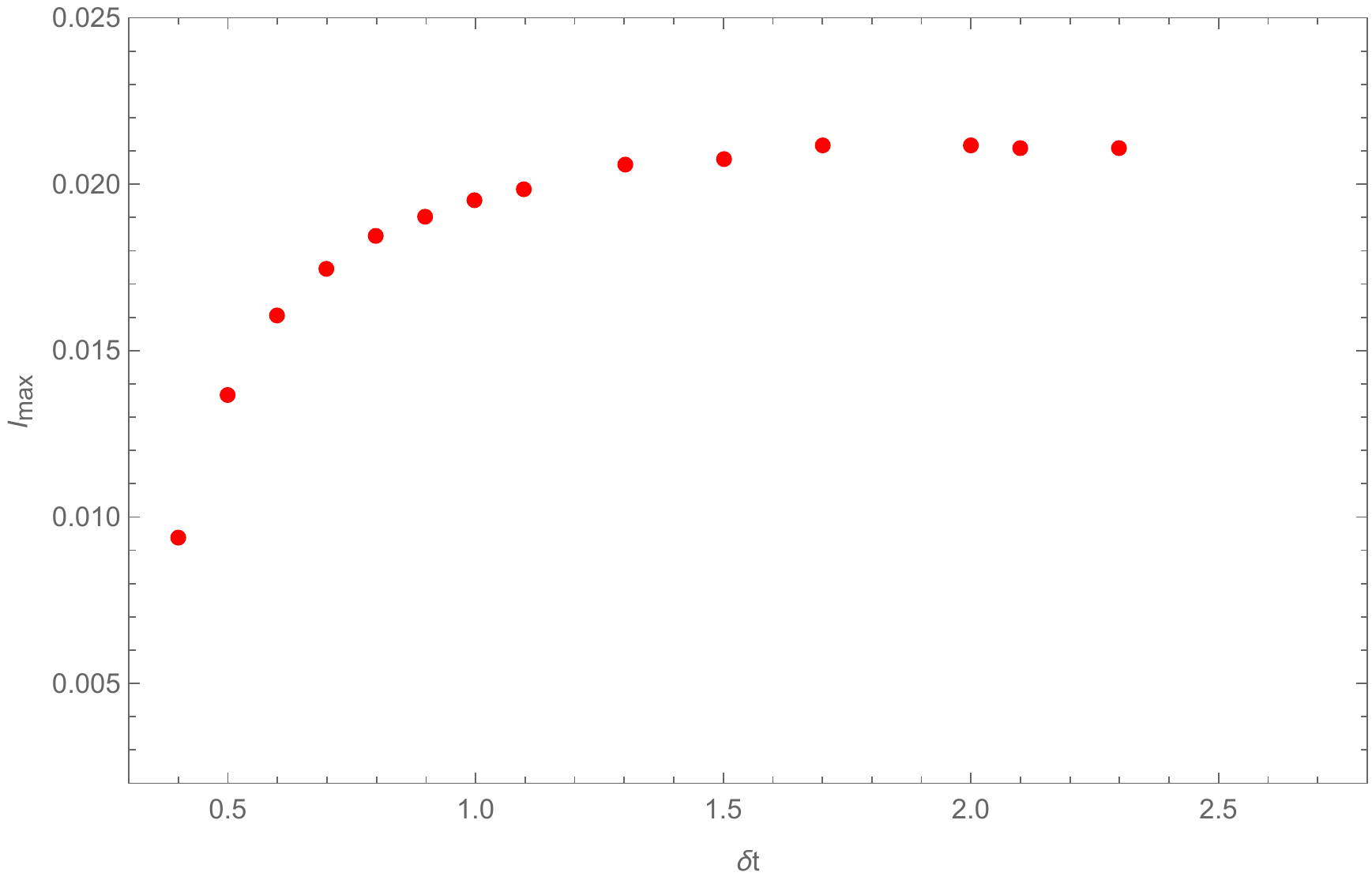} }
\caption{Top: The holographic mutual information $I$ as a function of the boundary time $t$ at fixed $\delta t=0.4$. In the left panel $l_{1}=1.4$ and $x=0.1$ and in the right one $l_{1}=1$ and $x=0.3$. The various curves correspond to different values of $l_{2}$. Some of the curves are invisible since everywhere disappearing such as $l_{2}=0.2$ (the purple curve).\\
Down: $Left$: The equilibrated value of mutual information as a function of $l_{2}$ at fixed $l_{1}=0.5$ and $x=0.2$. The different curves correspond to different value of $\delta t=2.5$(top) to $\delta t=0.4$(bottom). $Right$: Maximum value of equilibrated mutual information $I_{max}$ as a function of $\delta t$ for different values of $l_{2}=0.8,1.1,..,7.1,...10.1$ at fixed $l_{1}=0.5$, $x=0.2$.
}\label{fig2}
\end{figure}
Having studied the mutual information of the two sub-systems of the same length $l$, we shall extend the previous results to a situation where the two intervals have different lengths $l_{1}$ and $l_{2}$. As it was already mentioned, the mutual information for two sub-systems of the different lengths is given by
\begin{eqnarray}
I(l_1,l_2) = S(l_{1})+S(l_{2})-S(l_{1}\cup l_{2}).
\end{eqnarray}
In Fig. \ref{fig2}-top, we plot the time evolution of the holographic mutual information for two sub-systems $A$ and $B$, with lengths $l_{1}$ and $l_{2}$ respectively and separation length $x$, for two different regimes of $l_{2}$. It is significant to notice that there is an upper bound for $l_2$ due to (\ref{condition}) . On the left panel one can easily see that the mutual information increases as long as $l_2$ increases. It is intuitively comprehensible since for large $l_2$ the correlation of the two sub-systems $A$ and $B$ increases. On the other hand, for very large $l_2$ this correlation, or equivalently the mutual information, does not change substantially and it remains approximately constant. However, the same intuitive argument can not be applied to describe the behavior of the mutual information on the right panel. In fact, by increasing $l_2$ the final value of mutual information can be larger or smaller depending on the choice of $l_2$. As an example, if we consider $l_2=0.4, 0.9$ and $1.4$ we get $I(l_2=0.4)<I(l_2=0.9)>I(l_2=1.4)$ indicating that the mutual information has a maximum value, $I_{max}$. It is noticed that $\epsilon$ and $\delta t$ are kept fixed so that the final temperature, introduced in (\ref{temperature}), is the same for both left and right plots.
The (maximum) mutual information as a function of $l_{2}$ ($\delta t$) for different values of $\delta t$ ($l_2$) has been plotted in the left (right) panel of Fig. \ref{fig2}-down. In fact, the strategy is to change the final temperature of the system thus we consider various injection times. At low temperatures, corresponding to larger values of $\delta t$, the mutual information increases for large $l_2$ but it is not substantial for large enough $l_2$. By raising temperature, though the available $l_2$ is more limited according to (\ref{condition}), a maximum appears in the mutual information. Therefore, one can conclude that the final temperature and length $l_2$ have opposite effect on the mutual information which is in complete agreement with the results reported in \cite{Balasubramanian,Allais}.
It is noticed that the general results indeed coincide with the case $l_{1}=l_{2}$ so we merely express the following interesting outcomes. \\
\newpage
\begin{enumerate}
\item \textbf{Top plots:}
\begin{itemize}
\item In both plots if one increases the length of second sub-system $ l_{2}$, before the two sub-systems reach the final state, the two sub-systems become more and more entangled and hence the mutual information's peak goes upward. 
\item Remarkably, depending on the value of $l_{2}$ two different behaviors for the mutual information will be observed at the final state .
While, in the right plot increasing $l_{2}$ causes the mutual information decreases at the final state in the left plot if we increase the length of $l_{2}$, the mutual information will also increase. 
\item It is evident from the right plot that there is a length scale, $l_{2}\simeq0.3$, beyond which a disentangling transition occurs and then there is no correlation between the two sub-systems.
\end{itemize}
\item \textbf{Down plots:}
\begin{itemize}
\item The right panel indicates that if one decreases the final state temperature (corresponding to increase the injection time $\delta t$), the maximum value of the mutual information will increase gradually and then finally reaches an approximate constant value, $i.e.$ at low temperature regime $I_{max}$ stays constant. In other words, the higher the temperature of the final state is, the lesser the maximum value of the mutual information becomes.
\end{itemize}
\end{enumerate}
\begin{figure}[h] \label{fig:MI3}
\centering
{\includegraphics[width=0.4\textwidth]{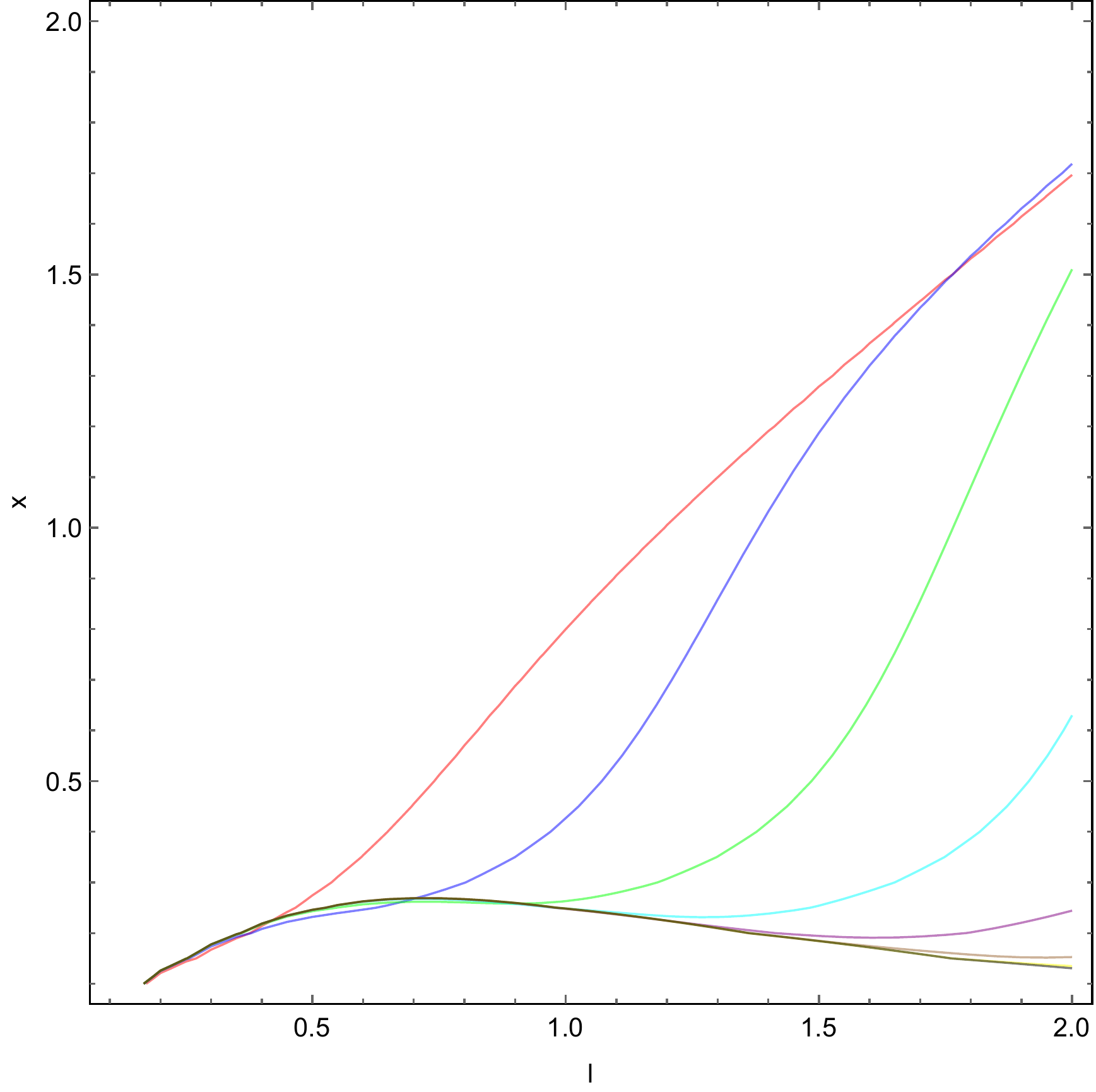} }
{\includegraphics[width=0.4\textwidth]{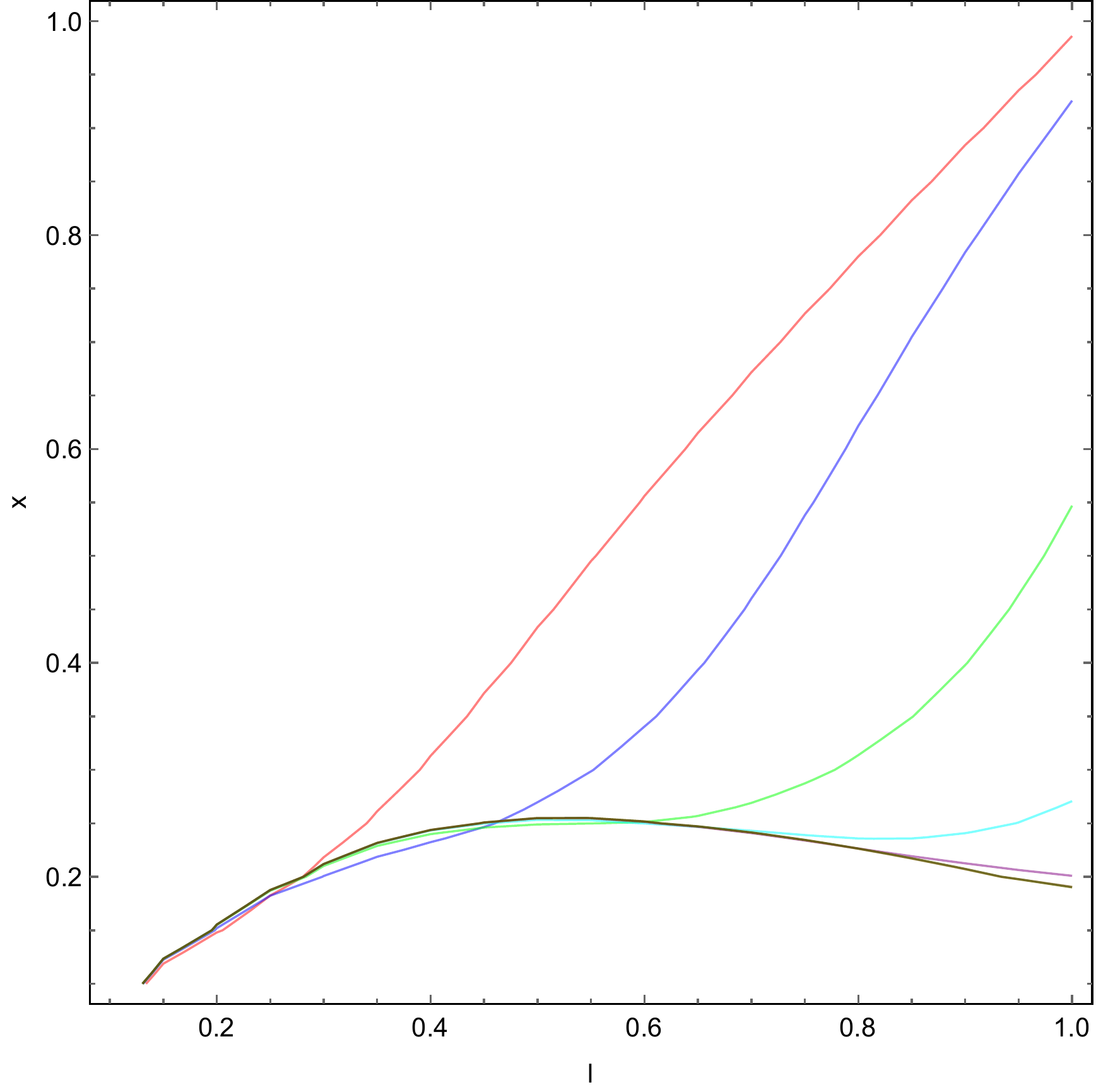} }
\caption{Two - dimensional parameter space of the two sub-systems living on the boundary theory. All of the curves correspond to $I=0$ (the transition curves) below which the two-subsystems become entangled. $Left$ : The transition curves of the two sub-systems, with the same length $l_{1}=l_{2}$, is plotted at fixed $\delta t=0.4 $ (for which $l , x < 2.1$) for different times $t=0.5$(red) to $t=4$(black). $Right$ : The transition curves of the two sub-systems with lengths $ l_{1}=l$ , $ l_{2}=2 l_{1}=2l $ is plotted at fixed $\delta t=0.4 $ (for which $l , x < 2.1$) for different times $t=0.5$(red) to $t=4$(black). }\label{fig3}
\end{figure}
In order to study the holographic mutual information more accurately we consider two sub-systems of the different length $l_1$ and $l_2$ separated by length $x$. Defining $H(t,l,x)\equiv S(t,2l+x)+S(t,x)-2S(t,l)$ we would like to find a family of curves in the configuration space, given by $x$ , $l$ and parameterized by $t$, satisfying the following equation
\begin{eqnarray}
H(t,l,x)=0,
\end{eqnarray}
corresponding to the time-dependent disentangling transition. In Fig. \ref{fig3}, this transition has been shown at different times for $l_{2}=l_{1}=l$ (left plot) and $l_{2}=2 l_{1}=2l$ (right plot). The area below each curve is a region where the two sub-systems have non-zero mutual information, $i.e.$ two sub-systems are entangled. In the following we list some interesting points regarding these two plots.
\begin{figure}[h] 
\centering
{\includegraphics[width=0.48\textwidth]{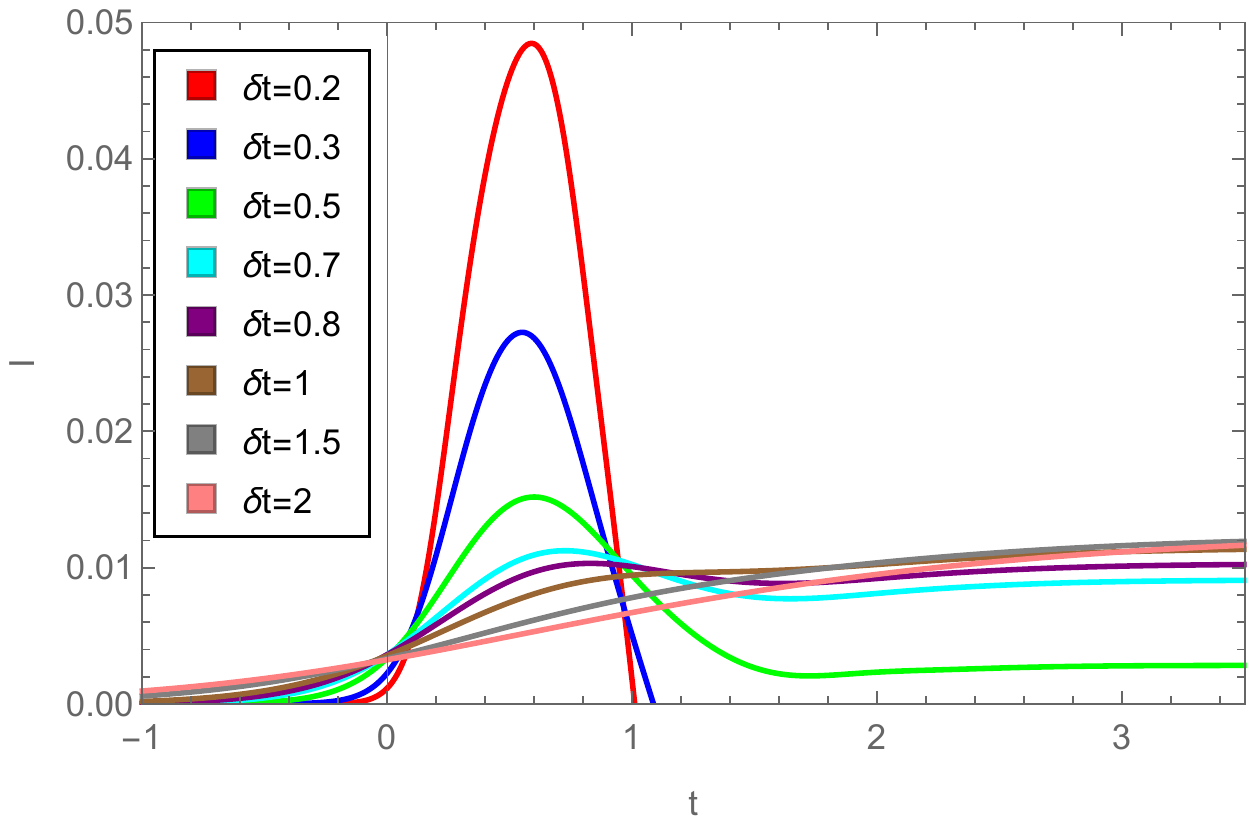} }
{\includegraphics[width=0.48\textwidth]{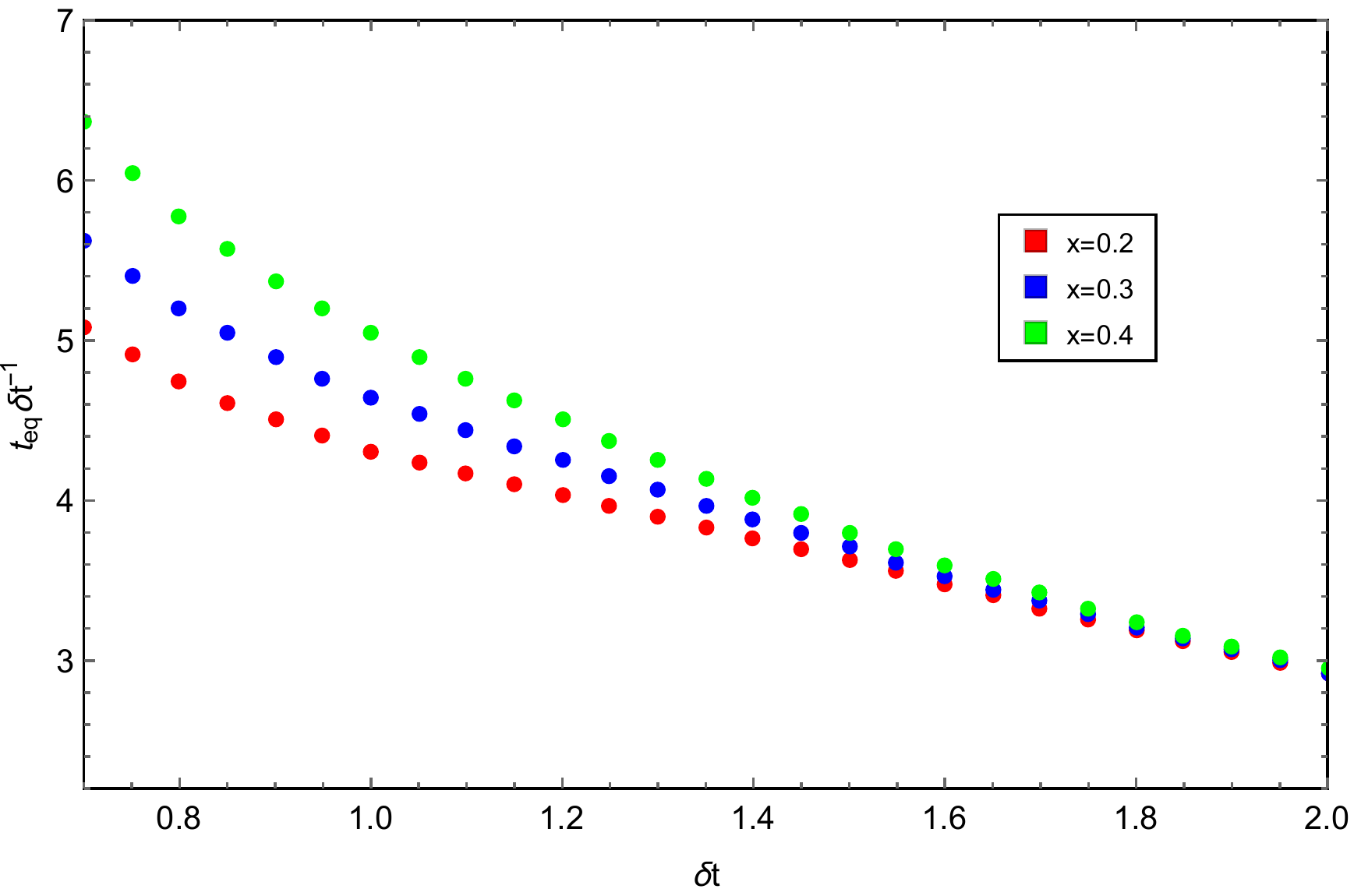} }
\caption{$Left$: The rescaled Holographic mutual information as a function of the boundary time $t$ at fixed $ l=1$ and $x=0.3$. The different curves are characterized by different values of $\delta t=0.2,0.3,0.4,0.6,0.9,1.5,2$. $Right$: The rescaled equilibrium time $t_{eq} \delta t ^{-1}$ as a function of the length $\delta t$ at fixed $ l=1$ for different separation length $x=0.2,0.3,0.4$. }\label{fig4}
\end{figure}
\begin{itemize}
\item One of the main features we observe is that there is a specific regime of the parameters, small enough $ l$ and $x$, where the mutual information is indeed independent of the time evolution. Hence one can say that transition curves do not feel the time lapse in the above regime.
\item Another interesting point is that, following the (\ref{EE}), configuration space is independent of the rate of the symmetry breaking, which specified by $\epsilon$ namely whatever $\epsilon$ is the transition curves will be the same as the previous one. Consequently, strength of the symmetry breaking has no role on the phase space of the two sub-systems.
\item Moreover, It can be observed from these plots that the region where the mutual information has non vanishing values in out-of-equilibrium time ($e.g.$ $t=0.5 $ top curve) is more wider than those of at the equilibrium time. In other words, there are wide region of parameters in out-of-equilibrium time where the two sub-systems become entangled. In fact, during the time evolution towards the final equilibrium state the phase space is more restricted.
\item Comparing the two plots we can clearly see that the qualitative features and behaviors are the same but it is worth to mention that in the case of $ l_{2}=2 l_{1} $ (right plot) there is an increase in the area of non-vanishing mutual information in the configuration space with respect to that of $l_{2}= l_{1} $ (left plot).
\end{itemize}
In Fig. \ref{fig4}, we plot the effect of quenching time $\delta t$ on the time evolution of the holographic mutual information for two sub-systems of the same length $l=1$ which are separated by $x=0.3$ (left plot) and the rescaled equilibration time $t_{eq} \delta t ^{-1}$ as a function of the quenching time $\delta t$ for different separation lengths $x$ (right plot). It can be seen from the left plot that although the quenching function $J(t)$ is a monotonically increasing function, the time evolution of the mutual information behaves in a different manner depending on the quenching rate $\delta t$. For fast enough quenches, $0.2\lesssim\delta t\lesssim0.8$ mutual information starts at the value roughly zero in the initial state and then reaches a peak in the intermediate times and finally declines to zero or constant values in the final state. While slower quenches $\delta t>0.8$ behave quite smoothly. Namely, if one increases the value of $\delta t$, there is no considerable gap for the mutual information between its initial state and its final state. Remarkably, we can observe that for slow quenches the mutual information approaches the adiabatic regime in the final state that is there is no dependence on the separation length $x$. Another interesting aspect noted from these plot is that this adiabatic behavior is completely independent of how large or small the symmetry of the initial state breaks. In fact, either the adiabatic behavior or the disentangling transition are indeed independent of the rate of the symmetry breaking.
\begin{figure}[h] 
\centering
{ \includegraphics[width=0.48\textwidth]{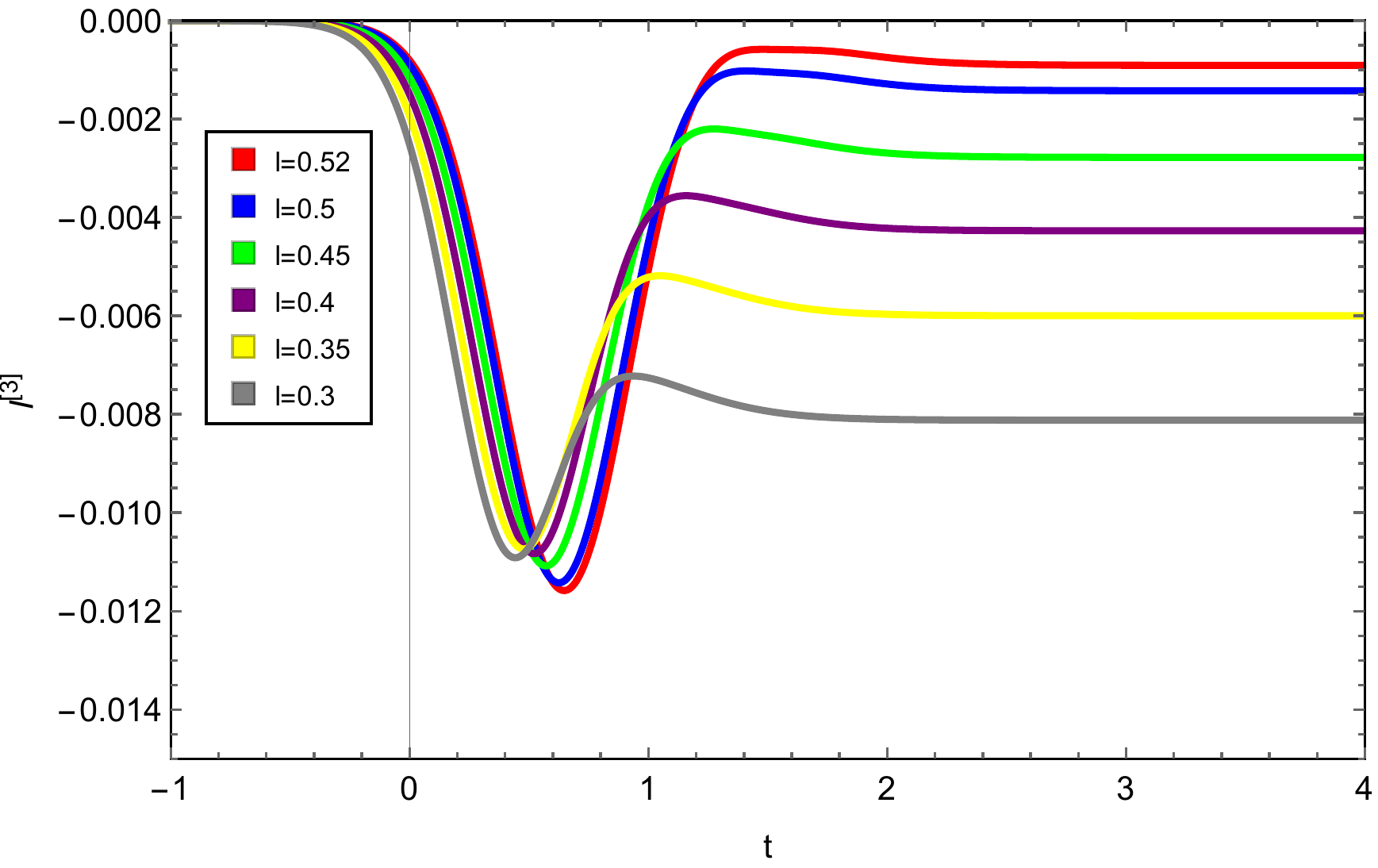} }
{\includegraphics[width=0.48\textwidth]{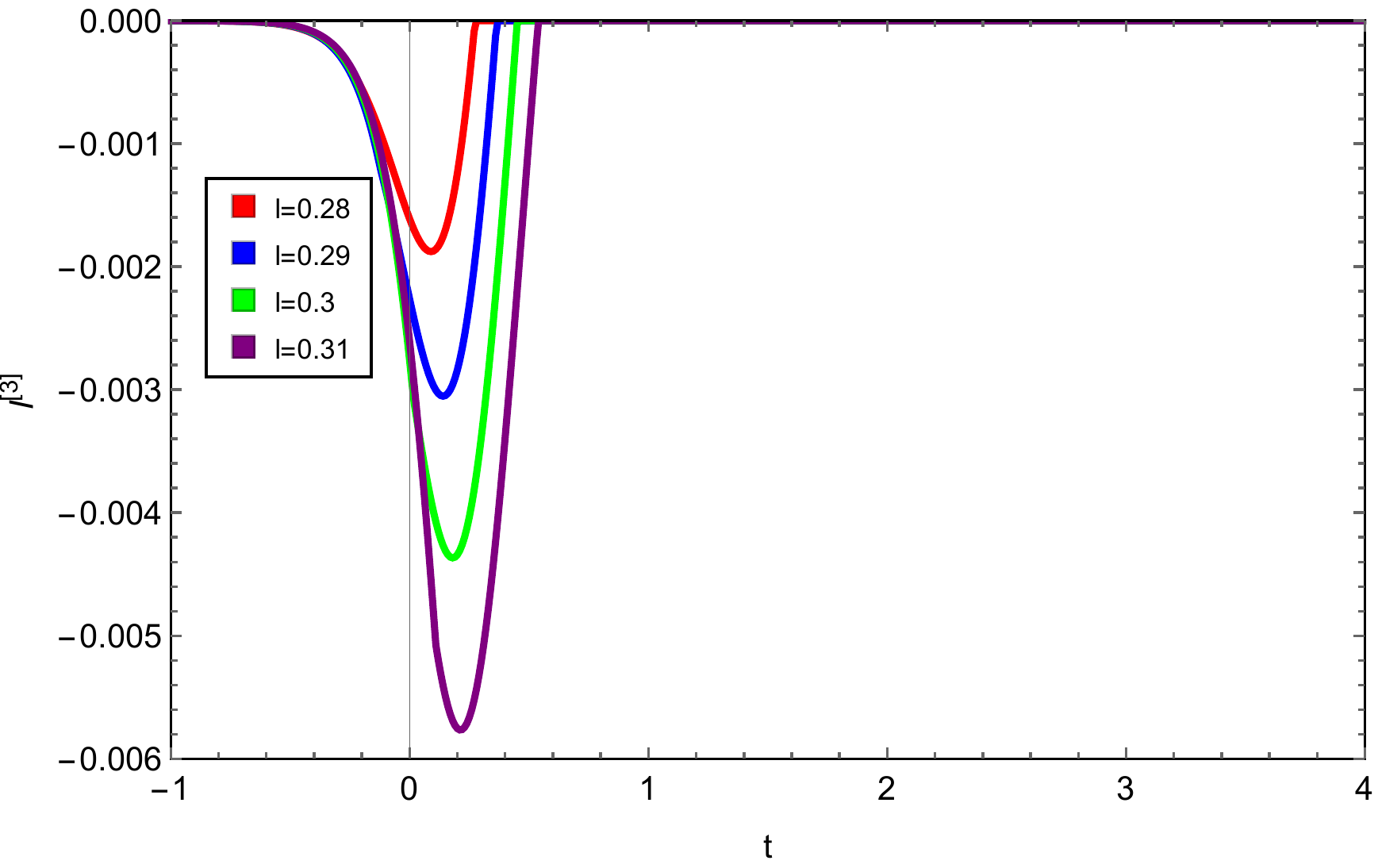} }
\caption{$Left$: The rescaled holographic tripartite information $I^{[3]}$ as a function of the boundary time $t$ at fixed $x=0.1$ and $\delta t=0.4 $ . The different curves are characterized by different values of $l=0.3,0.35,0.4,0.45,0.5,0.52$ which decrease from right to left (Some of the curves are not visible since everywhere vanishing). $Right$: The rescaled holographic tripartite information $I^{[3]}$ as a function of the boundary time $t$ at fixed $x=0.2$ and $\delta t=0.4 $. The different curves are characterized by different values of $l=0.28,0.29,0.3,0.31$ which decrease from right to left.} \label{fig5}
\end{figure} 
Now consider the time evolution of the tripartite information for three sub-systems of the same length $l$ with separation length $x$ living on the boundary. We plot in Fig. \ref{fig5} the results of the tripartite information for different values of $l$ and $x$ as a function of the boundary time.
The most important point is that the tripartite information is generically non-positive at all times, $i.e.$ the mutual information is extensive or superextensive. Hence, one can say that the holographic mutual information is monogamous which is coincided with \cite{Hayden}. Moreover, it is obvious that the tripartite information starts at the initial value (roughly zero) and ends at the final value, zero or more negative than the initial value, passing through an intermediate phase where it is absolutely negative. If one also decrease the length of the sub-systems, the tripartite information's peak tends to zero that is when the size of the sub-systems $l$ approaches the separation length $x$ the tripartite information becomes vanished. \\

\end{document}